\documentclass[aps,prd,reprint,superscriptaddress,amsmath,amssymb,nofootinbib,nolongbibliography,floatfix]{revtex4-2}

\usepackage{graphicx}
\usepackage{dcolumn}
\usepackage[utf8]{inputenc}
\usepackage{xcolor}
\usepackage[unicode=true,colorlinks,linkcolor=blue,anchorcolor=blue,urlcolor=blue,citecolor=blue,breaklinks=true]{hyperref}

\newcommand{\ii}{\mathrm{i}\,}

\newcommand{\pararrow}{\mathord{\buildrel{\lower3pt\hbox{$\scriptscriptstyle\leftrightarrow$}}\over {\partial}}} 
\newcommand{\pararrowk}[1]{\mathord{\buildrel{\lower3pt\hbox{$\scriptscriptstyle\leftrightarrow$}}\over {\partial}\hspace*{-0.18em}{}^#1}\hspace*{-0.18em} \,} 

\usepackage{physics}
\usepackage{bm}

\newcommand{\qfnu}{\affiliation{College of Physics and Engineering, Qufu Normal University, Qufu 273165, China}}

\newcommand{\ucas}{\affiliation{School of Physical Sciences, University of Chinese Academy of Sciences, Beijing 100049, China}}

\newcommand{\sucas}{\affiliation{Southern Center for Nuclear-Science Theory (SCNT), Institute of Modern Physics, Chinese Academy of Sciences, Huizhou 516000, Guangdong Province, China}}

\begin{document}

\title{Investigation of box diagrams for the peak of $Z_{cs}(3985)$ in $e^+e^- \to D^-_s D^{*0} K^+$ and $ D^{*-}_s D^0 K^+$}

\author{Yuan-Xin Zheng}\qfnu \ucas

\author{Gang Li}\email {gli@qfnu.edu.cn (Corresponding author)}  \qfnu
\author{Shi-Dong Liu} \qfnu
\author{Jia-Jun Wu}\email {wujiajun@ucas.ac.cn  (Corresponding author)}\ucas\sucas 

\begin{abstract}
The BESIII collaboration recently observed a charged hidden-charm structure with strangeness in the recoil mass spectrum of $K^+$ in the processes $e^+e^- \to D^-_s D^{*0} K^+$ or $D^{*-}_s D^0 K^+$, named as $Z_{cs}(3985)^-$. 
Within the energy region around the peak of $Z_{cs}(3985)$, various box diagrams are present. 
A systematic study of these box diagrams reveals the existence of a box singularity. 
Dalitz plots and invariant mass distributions of two charmed mesons from the box diagram for the processes $e^+e^- \to D^-_s D^{*0} K^+$ and $D^{*-}_s D^0 K^+$ are presented at five energy points $\sqrt s = 4.628,\,4.641,\,4.661,\,4.681$, and $4.698$ GeV. 
It is found that these box diagrams can generate peaks similar to those observed in experiments, suggesting that the peak of $Z_{cs}(3985)$ may be attributed to the box singularity contribution.
Additionally, slight differences in the invariant mass spectra of $D^-_s D^{*0}$ and $D^{*-}_s D^0$ are identified based on the box diagrams, needing further experimental validation.

\end{abstract}

\date{\today}







\maketitle

\section{Introduction}\label{sec:introduction}

Quantum chromodynamics (QCD) serves as the fundamental theory of strong interactions. 
In the low-energy regime, QCD exhibits non-perturbative characteristics that become significant at the hadronic level. 
The contributions of various hadronic loop diagrams, which must be taken into account alongside tree diagrams, are critical in this context. 
These hadronic loop diagrams involve loop integrals, and under certain conditions regarding the masses of the internal and external particles, singularities can arise, commonly referred to as Landau poles~\cite{Landau:1959fi}.
The presence of singularities can result in peak structures within the invariant mass spectrum of the final state particles. 
These peaks may be misidentified with those originating from resonant states, thus complicating the extraction of resonant parameters from experimental data. 
For example, the triangle singularity associated with triangle loop diagrams has implications for the interpretation of various peak structures, such as $a_1(1420)$ peak in the final state that is also explained as the effect from the triangle singularity in the $K^*\bar{K}K$ loop~\cite{COMPASS:2020yhb}.
To accurately assess the effects arising from these singular behaviors, a systematic analysis of loop diagram calculations within the relevant reactions is essential. 
A thorough understanding of the resulting line shapes is crucial for navigating the complexities these singularities introduced.

Since the discovery of $\chi_{c1}(3872)$ (or $X(3872)$ ) in 2003~\cite{Belle:2003nnu}, an increasing number of exotic states have been identified. 
Among these, certain hidden charm mesons exhibit quark flavor quantum numbers that clearly extend beyond those of traditional mesons formed by quark-antiquark pairs. 
Notable examples include the $Z_c(3900)$ mesons~\cite{BESIII:2013ris, Belle:2013yex} and the $Z_{cs}(3985/4000)$ mesons~\cite{BESIII:2020qkh, LHCb:2021uow}. 
Both $Z_c(3900)$ and $Z_{cs}(3985)$ are situated near the corresponding coupling channels $D\bar{D}^*$($+\bar{D}D^*$) and $D_s\bar{D}^*$ ($+\bar{D}_sD^*$), as well as $D_s^*\bar{D}$ ($+\bar{D}^*_s D$). 
Consequently, many researchers suggest that these states are likely hadronic molecular states~\cite{Yang:2020nrt, Meng:2020ihj, Baru:2021ddn, Yan:2021tcp, Sun:2020hjw, Du:2020vwb, Wang:2020htx, Dong:2020hxe, Guo:2020vmu}, please see various review papers~\cite{Guo:2017jvc, Chen:2016qju, Esposito:2016noz, Olsen:2017bmm, Karliner:2017qhf, Brambilla:2019esw, Meng:2022ozq}.
However, upon analyzing the one-boson exchange process, it becomes apparent that the $D_s^*\bar{D}$ and $D_s\bar{D}^*$ channels cannot exchange commonplace mesons such as $\rho$, $\omega$, $\pi$, or $K^{(*)}$. 
Instead, they can only exchange particles that contain both $q\bar{q}$($q=u,d$) and $s\bar{s}$ components, such as $\eta$ and $\eta'$, or some heavy mesons. 
Analytical consideration of light flavor quark lines indicates that this boson exchange in the $t$-channel is indeed disconnected in terms of light flavor quark lines, which is fundamentally different from the situation in $D\bar{D}^*$ scenario. 
Therefore, in general molecular state analyses, it is found to be challenging to form bound states of $D_s^*\bar{D}$ and $D_s\bar{D}^*$.
Similar results are also found in Refs.~\cite{Chen:2020yvq, Ikeno:2020mra, Ortega:2021enc}, such $Z_{cs}(3985/4000)$ can not be an resonance or bound states of $D_s^*\bar{D}$ and $D_s\bar{D}^*$ channels.
Thus, the $Z_{cs}$ particles observed by the BESIII collaboration and the LHCb  collaboration require scrutiny to determine whether they genuinely exist. 
Notably, Refs.~\cite{Wang:2020kej, Ikeno:2020mra, Llanes-Estrada:2021jud, Abreu:2022jmi} suggested that $Z_{cs}(3985)$ may be formed through a kinematic effect caused by the loop diagrams. 
However, these analyses often overlook the contributions from box diagrams, which is the focal point of the present study.

The study of singularity behavior dates back to 1959 with Landau's work~\cite{Landau:1959fi}, followed by a series of related investigations over the subsequent decades~\cite{Mathews:1959zz, Coleman:1965xm, Schmid:1967ojm}. 
However, due to limitations in experimental techniques at that time, it was challenging to meet the stringent kinematic conditions required for these investigations, although they try to observe such triangle singularity in various processes~\cite{Peierls:1961zz, Chang:1964fgx, Goebel:1964zz, Aitchison:1964zz, Goebel:1982yb, He:1984ev}. 
It was not until 2012 that the BESIII collaboration observed the isospin-breaking process $\eta_c(1405) \to 3\pi$, revealing an extremely narrow contribution from $f_0(980)$ resonance and an isospin-breaking strength that exceeded typical expectations~\cite{BESIII:2012aa}. 
In response to this finding, theoretical literatures~\cite{Wu:2011yx, Wu:2012pg, Aceti:2012dj, Du:2019idk} introduced the concept of triangle singularities to explain this phenomenon. 
Since then, triangle singularities have re-emerged in the research landscape, prompting investigations in various related reactions, as discussed in recent review paper~\cite{Guo:2019twa} and subsequent studies~\cite{Hsiao:2019ait, Nakamura:2019nwd, Sakai:2020ucu, Sakai:2020fjh, Molina:2020kyu, Braaten:2020iye, Shen:2020gpw, Abreu:2020jsl, Huang:2020kxf, Ikeno:2021frl, Feijoo:2021jtr, Molina:2021awn, Huang:2021olv, Luo:2021hyy, Achasov:2022onn, Du:2022nno, Wang:2022wdm, Burns:2022uiv, Meng:2022wgl, Yan:2022eiy, He:2023plt, Duan:2023dky, Wang:2023xua, Bayar:2023azy, Xiao:2024ohf, Huang:2024oai, Sakthivasan:2024uwd, Wang:2024ewe, Yu:2024sqv, Huang:2025rvj}. Very recently, Duan $et$ $al$. argued  that there exist the enhancements at the thresholds induced by the triangle and box singularities~\cite{Duan:2023dky};
Triangle singularities represent a specific type of loop diagram singularity; there are also studies~\cite{Nakamura:2021bvs, Nakamura:2021qvy} addressing the singularities of double-loop diagrams. 
Moreover, within single-loop diagrams, numerous other higher-order divergences exist, such as the box singularity arising from box diagrams. 
Notably, in the processes $e^+ e^- \to D_s^{-} D^{*0} K^+$ and $e^+ e^- \to D^{*-}_s D^0 K^+$, the potential contributions from box diagrams have not yet been investigated.
%
Recently, literature~\cite{Shen:2025nen} has outlined the kinematic conditions under which box diagrams can exhibit singularities. 
In this paper, we will systematically analyze the possible contributions of box diagrams in these two processes.

The structure of the paper is as follows: Sec.~\ref{sec:formula} introduces the theoretical calculations and formulas related to loop diagrams. 
Sec.~\ref{sec:results} presents the calculation results and related discussions. Finally, we conclude with a summary and outlook in Sec.~\ref{sec:summary}.


\section{Theoretical Framework}\label{sec:formula}

\subsection{Box Diagrams and Effective Lagrangians}
\label{subsec:effective Lagrangians}
\begin{figure*}
    \centering
    \includegraphics[width=0.9\hsize]{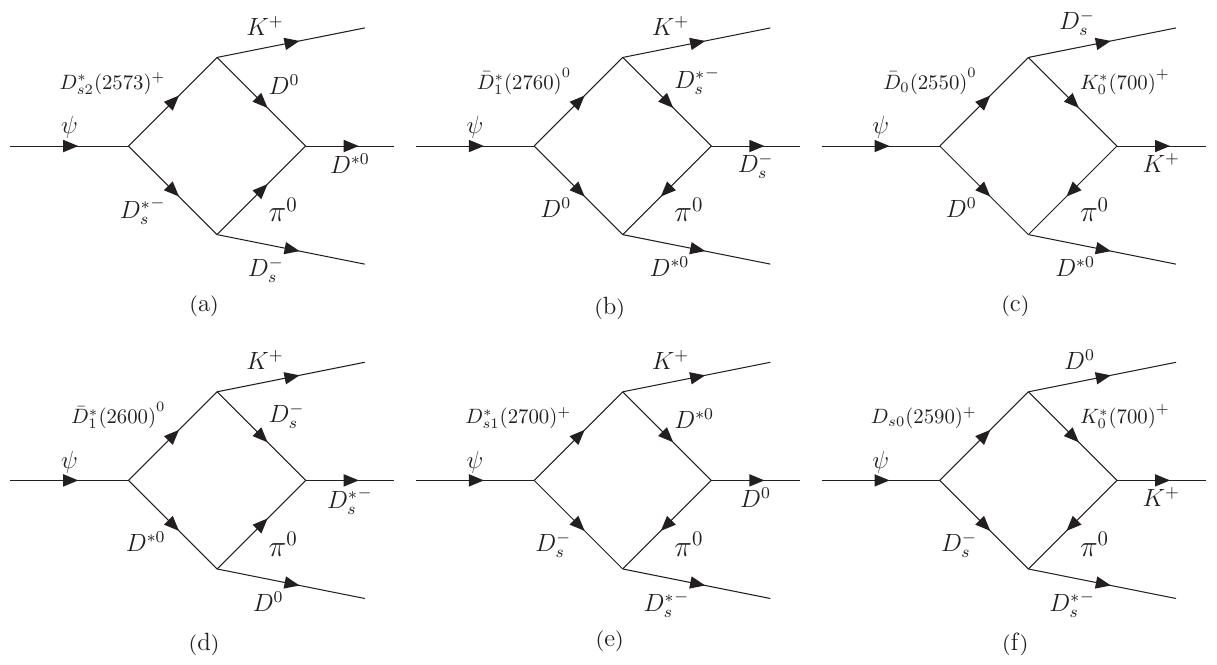}
    \caption{Feynman diagrams for the $\psi \to D^-_s D^{*0} K^+$ [(a)-(c)] and $D^{*-}_s D^0 K^+$ [(d)-(f)] via box loops.}\label{fig:boxtu}
\end{figure*}

In this study, we investigate the processes $e^+e^- \to D^-_s D^{*0} K^+$ and $e^+e^- \to D^{*-}_s D^0 K^+$ by considering the intermediate charmoniumlike resonances $\psi$, which decay into the final states via the box loops shown in Fig.~\ref{fig:boxtu}. 
As seen, the diagrams (a)-(c) contribute to the process $\psi \to D^-_s D^{*0} K^+$, while the diagrams (d)-(f) depict the $\psi \to D^{*-}_s D^0 K^+$.

As described in reference~\cite{Shen:2025nen}, box diagrams can be categorized into two main types. 
The first type involves two particles decayed from the parent particle, with each of these particles subsequently undergoing decay.
Their respective decay products eventually collide to form the final state particles, as illustrated in Figs.~\ref{fig:boxtu}(a) and (d).
The second type involves one particle from the initial decay of the parent particle undergoing two additional decays, ultimately interacting with another decay particle of the parent, as depicted in Figs.~\ref{fig:boxtu}(b, c) and (e, f). 
In the reactions $\psi \to  D^{-}_s D^{*0} K^{+}$ and $\psi \to  D^{*-}_s D^{0} K^{+}$, the final states include only $D^{*0}$ and $D^{*-}_s$, which may arise from particle collisions rather than being produced through decays. 
Consequently, Fig.~\ref{fig:boxtu} encompasses virtually all the mechanisms that could contribute to the occurrence of box diagram singularities. 
Furthermore, if we do not assume existence of a bare $Z_{cs}(3895)$, the triangle loop effect will have a four-particle vertex since there are three particles in the final states. However, such four-particle vertex includes an assumption of two three-particle vertices with an intermediate state. In principle, our box diagrams have already included such contribution.

Here, for the intermediate particles, at the first vertex we consider $D^*_{s2}(2573)(J^P = 2^+)$, ${\bar D}^*_1(2760)$, ${\bar D}_0(2550)$, ${\bar D}^*_1(2600)$, $D^*_{s1}(2700)$, and $D_{s0}(2590)$, whose parameters are taken from Ref.~\cite{ParticleDataGroup:2024cfk}.
These selected resonances ensure that all particles in the box loop could be on their mass shells, thereby allowing for the occurrence of box singularities.
The singularity would result in a peak in the $D^-_s D^{*0}$ and $D^{*-}_s D^0$ invariant mass spectra~\cite{Shen:2025nen}.
Using the condition equations for box singularity, we find these diagrams would produce the box singularity. However, the sharp peak maybe missing since these $D$ and $D_s$ resonances have large decay widths.

The interactions of the $S$-wave charmonium $\psi$ with the $D(0^-)$ and $D^*(1^-)$ mesons are described by the Lagrangian ~\cite{Oh:2000qr,Colangelo:2003sa}
\begin{eqnarray}\label{eq:Lpsi}
    	\mathcal{L}_S &=& \ii g_{\psi DD} \psi_\mu \bar{D} \pararrowk{\mu}D \nonumber\\
    	&+& g_{\psi D^*D}\epsilon_{\mu\nu\alpha\beta}\partial^\mu \psi^{\nu} (D^{*\alpha} \pararrowk{\beta}\bar{D} - D \pararrowk{\beta}\bar{D}^{*\alpha}) \nonumber\\
    	& -&\ii g_{\psi D^*D^*} \psi_\mu (D^*_\nu \pararrowk{\nu}\bar{D}^{*\mu} + D^{*\mu}\pararrowk{\nu}\bar{D}^*_\nu \nonumber\\
    	&-& D^*_\nu \pararrowk{\mu}\bar{D}^{*\nu})+ \mathrm{H.c.}\,,
\end{eqnarray}
where the $D$ and $D^*$ stand for the pseudoscalar and vector charmed mesons considered in Fig.~\ref{fig:boxtu}, respectively. 
The $\bar{D}$ and $\bar{D}^*$ are the corresponding anticharmed ones. 
The form of the relevant coupling coefficient used is as follows~\cite{Wu:2021ezz,Zheng:2024eia}
\begin{subequations}\label{eq:gpsidds}
\begin{align}
    g_{\psi\bar{D}D} &= 2g_1 m_D \sqrt{m_\psi}\,,\label{eq:gpsidd}\\
    g_{\psi D^*D}&= 2g_1 \sqrt{m_D m_{D^*}/m_\psi}\,,\label{eq:gpsidsd}\\
    g_{\psi \bar{D}^*D^*}& = 2g_1 m_{D^*}\sqrt{m_\psi}\,.\label{eq:gpsidsds}
\end{align}
\end{subequations}
Here the values of $m_\psi$ are taken to be the center-of-mass energies of interest.

The interaction of the $\psi$ with the $D^*_{s2}(J^P = 2^+)$ and charmed meson is described by~\cite{Zheng:2024eia}
\begin{equation}\label{eq:Ds2Dspsi}
    	\mathcal{L}_{\psi \bar{D}^*_{s2} D^*_s}  =  g_{\psi \bar{D}^*_{s2} D^*_s}{\psi_\mu}D^{*+\mu\nu}_{s2}D^{*-}_{s\nu} + \mathrm{H.c.}\,.
\end{equation}
In addition, the Lagrangian of $D^*_{s2}$ with charmed meson and pseudoscalar kaon reads \cite{Huang:2021kfm}
\begin{eqnarray}\label{eq:Ds2PP}
    	\mathcal{L}_{D^*_{s2} D K} &=& -g_{D^*_{s2} D K}D^{*\mu\nu}_{s2} \partial_\mu D\partial_\nu K^\dagger + \mathrm{H.c.}\,.
\end{eqnarray}

 Based on the heavy quark limit and chiral symmetry, the interactions between the Goldstone pseudoscalar and charmed mesons can be described by the Lagrangian~\cite{Wu:2021udi, Wang:2022qxe, Wu:2022hck, Zheng:2024eia},
\begin{eqnarray}\label{eq:LP}
    	\mathcal{L} &=& -\ii g_{D^*D\mathcal{P}}\big(D^{i \dagger}\partial^{\mu} \mathcal{P}_{ij}^\dagger D_\mu^{*j} - D_\mu^{*i\dagger}\partial^\mu \mathcal{P}_{ij}^\dagger D^j\big) \nonumber\\
    	&+& \frac{1}{2}g_{D^*D^*\mathcal{P}}\epsilon_{\mu\nu\alpha\beta} D_i^{*\mu\dagger}\partial^\nu
    	\mathcal{P}^{ij\dagger}\pararrowk{\alpha} D_j^{*\beta} + \mathrm{H.c.}\,,
\end{eqnarray}
where the $\mathcal{P}$ is $3\times 3$ pseudoscalar mesons in the matrix form
\begin{equation}\label{eq:pmatrix}
    \mathcal{P} = 
    \begin{pmatrix}
    \frac{\pi^0}{\sqrt{2}} + \frac{\eta}{\sqrt{6}} & \pi^+ & K^+\\
		\pi^- & -\frac{\pi^0}{\sqrt{2}} + \frac{\eta}{\sqrt{6}} & K^0\\
		K^- & \bar{K}^0 & -\sqrt{\frac{2}{3}}\eta
    \end{pmatrix}\,.
\end{equation}
The relevant coupling constants are
determined by using the following relations~\cite{Wu:2021udi, Zheng:2024eia}
\begin{equation}\label{eq:gddps}
    g_{D^*D^*\mathcal{P}} = \frac{g_{D^*D\mathcal{P}}}{\sqrt{m_D m_{D^*}}}= \frac{2g}{f_\pi}\,.
\end{equation}

The isospin-violating process $D^*_s \to D_s \pi^0$ results from the $\pi^0$-$\eta$ mixing. 
Therefore, this transition is studied in an effective Lagrangian couplings as~\cite{Wu:2012pg,Wu:2011yx}
	\begin{align}\label{eq:VPP}
	\mathcal{L}_{D^*_s D_s \pi^0} &= g_{D^*_s D_s \pi^0} D_s \partial_\mu \pi^0 D^{*\mu}_s \,,
   	\end{align}
where $g_{D^*_s D_s \pi^0}$ is an effective coupling constant including the isospin violation factor.

For the vertices of the light scalar and pseudoscalar mesons, we take the Lagrangian~\cite{Wu:2012pg,Wu:2011yx}
\begin{equation}\label{eq:SPP}
   \mathcal{L}_{S P P} = g_{D_s D\kappa} D_s D\kappa + g_{K\kappa \pi}K\kappa \pi
\end{equation}
with $g_{D_s D\kappa}$ and $g_{K\kappa \pi}$ being the coupling constants.

\subsection{Amplitudes of $\psi \to D^-_s D^{*0} K^+$ and $D^{*-}_s D^0 K^+$}

According to the Lagrangians above, we can obtain the amplitudes for $\psi(p) \to K^+(p_1) D^{*0}(p_2) D^-_s(p_3)$ corresponding to Fig.~\ref{fig:boxtu}(a)-(c):
	\begin{align}
		\mathcal{M}^a_{\mu\beta} & = \int \frac{\mathrm{d}^4q}{(2\pi)^4}\big[g_{\psi \bar{D}^*_{s2} D^*_s}\,g_{\mu\alpha} g_{\rho\sigma}\big]\big[g_{D^*_{s2} D^*_s K}\,p_{1\delta} q_{4\gamma}\big]\nonumber\\
		&\times \big[-\frac{1}{\sqrt 2} g_{D^* D \mathcal{P}} \,q_{3\beta}\big]\big[g_{D^*_s D_s \pi^0}\,p_{3\nu}\big]\nonumber\\
		&\times S^{\alpha\sigma\delta\gamma}(q_1,m_{D^*_{s2}})S^{\rho\nu}(q_2,m_{D^*_s})\nonumber\\
		&\times S(q_3,m_\pi)S(q_4,m_{D^0})\,,
	\end{align}
	\begin{align}
		\mathcal{M}^b_{\mu\beta} & = \int \frac{\mathrm{d}^4q}{(2\pi)^4}\big[- g_{\psi \bar{D}^* D}\epsilon_{\nu\mu\alpha\gamma}\,p^\nu(q_1 - q_2)^\gamma \big]\nonumber\\
		&\times \big[\frac{1}{2}g_{D^* D^* \mathcal{P}}\epsilon_{\rho\sigma\delta\xi}\,p_1^\sigma (q_1 + q_4)^\delta \big]\big[- g_{D^*_s D_s \pi^0}\, q_{3\theta}\big]\nonumber\\
		&\times \big[- \frac{1}{\sqrt{2}} g_{D^* D \mathcal{P}}\,q_{3\beta}\big] S^{\alpha\xi} (q_1,m_{\bar{D}^*_1(2760)})\nonumber\\
		&\times S(q_2,m_{D^0}) S(q_3,m_\pi) S^{\rho\theta}(q_4,m_{D^*_s}) \,,
	\end{align}
	\begin{align}
		\mathcal{M}^c_{\mu\beta} & = \int \frac{\mathrm{d}^4q}{(2\pi)^4}\big[g_{\psi \bar{D} D}(q_1 - q_2)_\mu \big]\big[g_{D_s D \kappa}\big]\big[g_{K \kappa \pi}\big]\nonumber\\
		&\times \big[- \frac{1}{\sqrt{2}}g_{D^* D\mathcal{P}}\,q_{3\beta}\big] S(q_1,m_{\bar{D}_0(2550)})\nonumber\\
		&\times S(q_2,m_{D^0})S(q_3,m_\pi)S(q_4,m_\kappa)\,.
	\end{align}

For the $\psi(p) \to K^+(p_1) D^{*-}_s(p_2) D^0(p_3)$, the amplitudes due to the diagrams in Fig.~\ref{fig:boxtu}(d)-(f) are 
	\begin{align}
		\mathcal{M}^d_{\mu\beta} & = \int \frac{\mathrm{d}^4q}{(2\pi)^4}\big[\ii g_{\psi \bar{D}^* D}((q_2 - q_1)_\alpha g_{\mu\gamma}\nonumber\\
		& +(q_2 - q_1)_\gamma g_{\mu\alpha}
		 - (q_2 - q_1)_\mu g_{\gamma\alpha})\big]\nonumber\\
		& \times \big[g_{D^* D\mathcal{P}}\, p_{1\sigma}\big]\big[g_{D^*_s D_s \pi^0}\,q_{4\beta}\big] \big[\frac{1}{\sqrt{2}}g_{D^* D\mathcal{P}}\,p_{3\delta}\big]\nonumber\\
		& \times S^{\alpha\sigma}(q_1,m_{\bar{D}^*_1(2600)^0}) S^{\gamma\delta}(q_2,m_{D^0})\nonumber\\
		& \times S(q_3,m_\pi) S(q_4,m_{D_s})\,,
	\end{align}
	\begin{align}
		\mathcal{M}^e_{\mu\beta} & = \int \frac{\mathrm{d}^4q}{(2\pi)^4}\big[- g_{\psi \bar{D}^* D}\epsilon_{\nu\mu\alpha\gamma}\,p^\nu(q_1 - q_2)^\gamma \big]\nonumber\\
		& \times \big[\frac{1}{2}g_{D^* D^* \mathcal{P}}\epsilon_{\rho\sigma\delta\xi}\,p_1^\sigma(q_1 + q_4)^\delta \big]\big[g_{D^*_s D_s \pi^0}\,q_{2\beta}\big]\nonumber\\
		& \times \big[- \frac{1}{\sqrt{2}} g_{D^* D\mathcal{P}}\,q_{3\theta}\big] S^{\alpha\xi}(q_1,m_{D^*_{s1}})\nonumber\\
		& \times S(q_2,m_{D_s}) S(q_3,m_\pi) S^{\rho\theta}(q_4,m_{D^{*0}})\,,
	\end{align}
	\begin{align}
		\mathcal{M}^f_{\mu\beta} & = \int \frac{\mathrm{d}^4q}{(2\pi)^4}\big[g_{\psi \bar{D} D}(q_1 - q_2)_\mu \big]\big[g_{D_s D\kappa}\big]\big[g_{K \kappa \pi}\big]\nonumber\\
		& \times \big[- \frac{1}{\sqrt{2}}g_{D^* D\mathcal{P}}\,q_{3\beta}\big] S(q_1,m_{D_{s0}(2590)})\nonumber\\
		& \times S(q_2,m_{D_s}) S(q_3,m_\pi) S(q_4,m_\kappa)\,.
	\end{align}
Here the $S$, $S^{\mu\nu}$, and $S^{\mu\nu\alpha\beta}$ represent the propagators for the charmed mesons $D$, $D^*$, and $D^*_{s2}$, respectively. Their forms can be written as \cite{Zou:2002ar}
\begin{subequations}
	\begin{align}
		S(q,m) &= \frac{1}{q^2 - m^2 + \ii m \Gamma}\,,\\
		S^{\mu\nu}(q,m) &= \frac{-g^{\mu\nu}+q^\mu q^\nu /m^2}{q^2 - m^2 + \ii m \Gamma}\,,\\
		S^{\mu\nu\alpha\beta}(q,m) &= \frac{P^{(2)}_{\mu\nu\alpha\beta}(q)}{q^2 - m^2 + \ii m \Gamma}\,.
	\end{align}
\end{subequations}
Here $P^{(2)}_{\mu\nu\alpha\beta}(q)$, the projection operator for the $D_{s2}^\ast$, is
\begin{align}
        P^{(2)}_{\mu\nu\alpha\beta}(q) &= \frac{1}{2}[\tilde{g}^{\mu\alpha}(q)\tilde{g}^{\nu\beta}(q) + \tilde{g}^{\mu\beta}(q)\tilde{g}^{\nu\alpha}(q)] \nonumber \\ &  - \frac{1}{3} \tilde{g}^{\mu\nu}(q)\tilde{g}^{\alpha\beta}(q)  
\end{align}
with
\begin{equation}
		\tilde{g}^{\mu\nu}(q)=g^{\mu\nu}-\frac{q^\mu q^\nu }{q^2}\,.
\end{equation}

The differential decay width of the $\psi \to D^-_s D^{*0} K^+(D^{*-}_s D^0 K^+)$ is described by
\begin{equation}\label{eq:width}
    	\mathrm{d}\Gamma_{\mathrm{tot}} = \frac{1}{3}\frac{1}{64{\pi^3}{m^3_\psi}} m_{13} m_{23} \overline{\left|\mathcal{M}_{\mathrm{tot}}\right|^2} \mathrm{d}m_{13} \mathrm{d}m_{23}\,.
\end{equation}
Here, the overline indicates the summations over the spins of initial and final states; $m_{13}$ and $m_{23}$ are the invariant masses of the final states, defined as $m^2_{13}=(p_1+p_3)^2$ and $m^2_{23}=(p_2+p_3)^2$, respectively \cite{ParticleDataGroup:2024cfk}.

\section{Numerical Results and Discussion}\label{sec:results}

\subsection{Line shapes of $\psi \to D^-_s D^{*0} K^+(D^{*-}_s D^0 K^+)$ processes}

The amplitude formulas related to the box diagrams in Fig~\ref{fig:boxtu} have been provided in Sec.~\ref{sec:formula}. 
However, the interaction strength of certain vertices cannot be precisely determined, such as the isospin-breaking vertex of $D^*_s D_s \pi^0$. 
Therefore, in this calculation, we can only present the contributions corresponding to the box diagram as a linear function of the invariant mass of the final state particles at fixed energy, along with the corresponding Dalitz plot. 
The relevant coupling constants are temporarily set to 1.

\begin{figure*}
	\centering
	\includegraphics[width=0.95\linewidth]{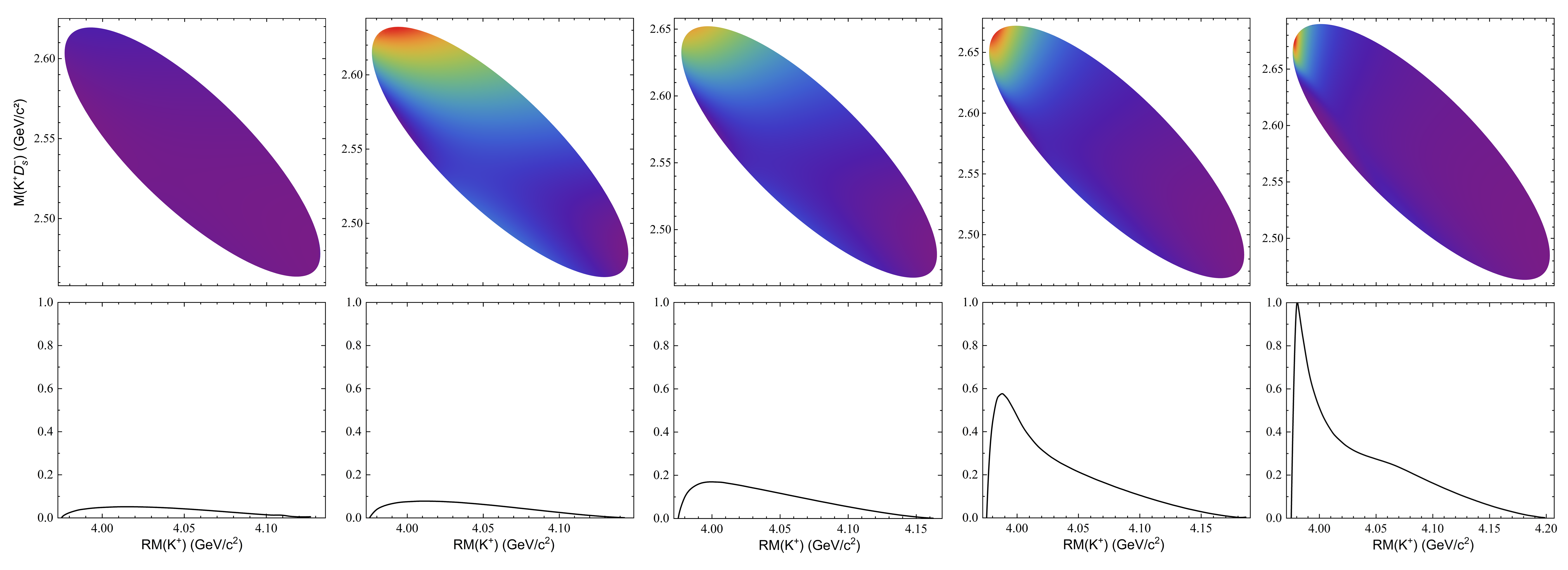}
	\caption{Dalitz plots and invariant mass spectra of the process $\psi \to D^-_s D^{*0} K^+$ in Fig.~\ref{fig:boxtu}(a). The $\sqrt s$ from left to right is $4.628,\,4.641,\,4.661,\,4.681$ and $4.698$ GeV. It should be noticed that in the invariant mass spectra the maximum of the value has been normalized to 1.0\,.}
	\label{fig:Invariantmassspectra1}
\end{figure*}

\begin{figure*}
	\centering
	\includegraphics[width=0.95\linewidth]{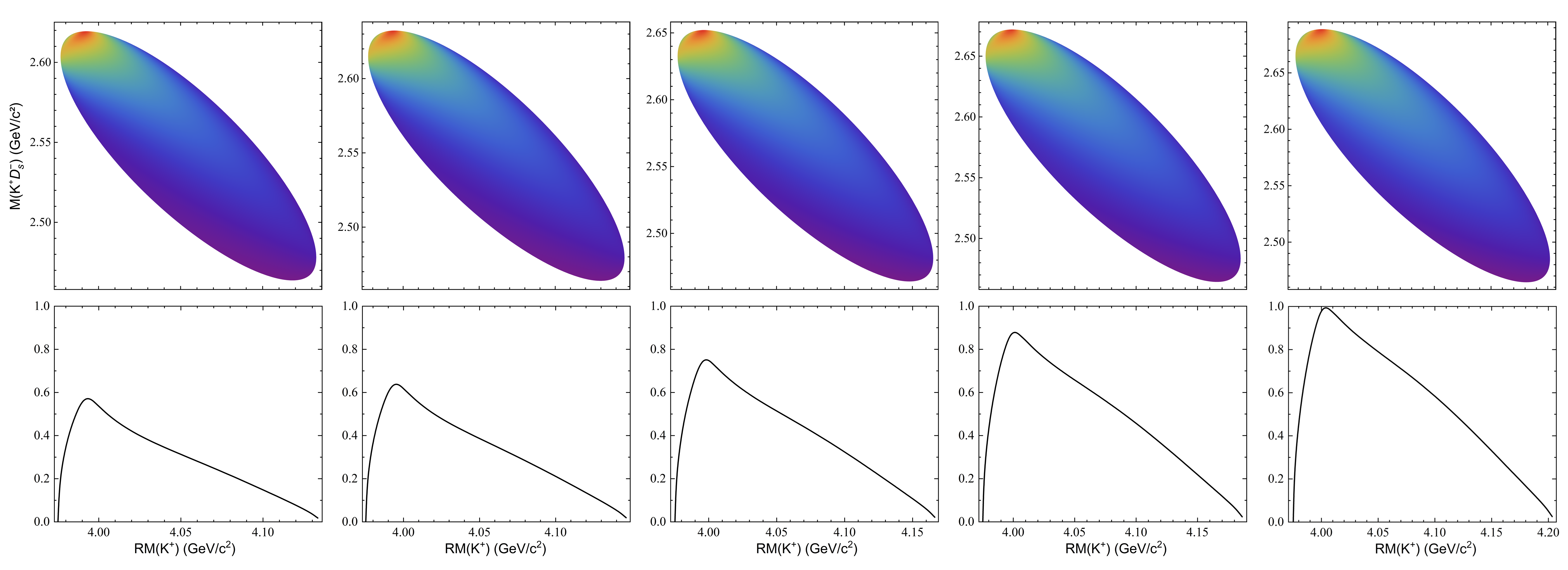}
	\caption{Dalitz plots and invariant mass spectra of the process $\psi \to D^-_s D^{*0} K^+$ in Fig.~\ref{fig:boxtu}(b). Further caption text as in \ref{fig:Invariantmassspectra1}.}
	\label{fig:Invariantmassspectra2}
\end{figure*}

\begin{figure*}
	\centering
	\includegraphics[width=0.95\linewidth]{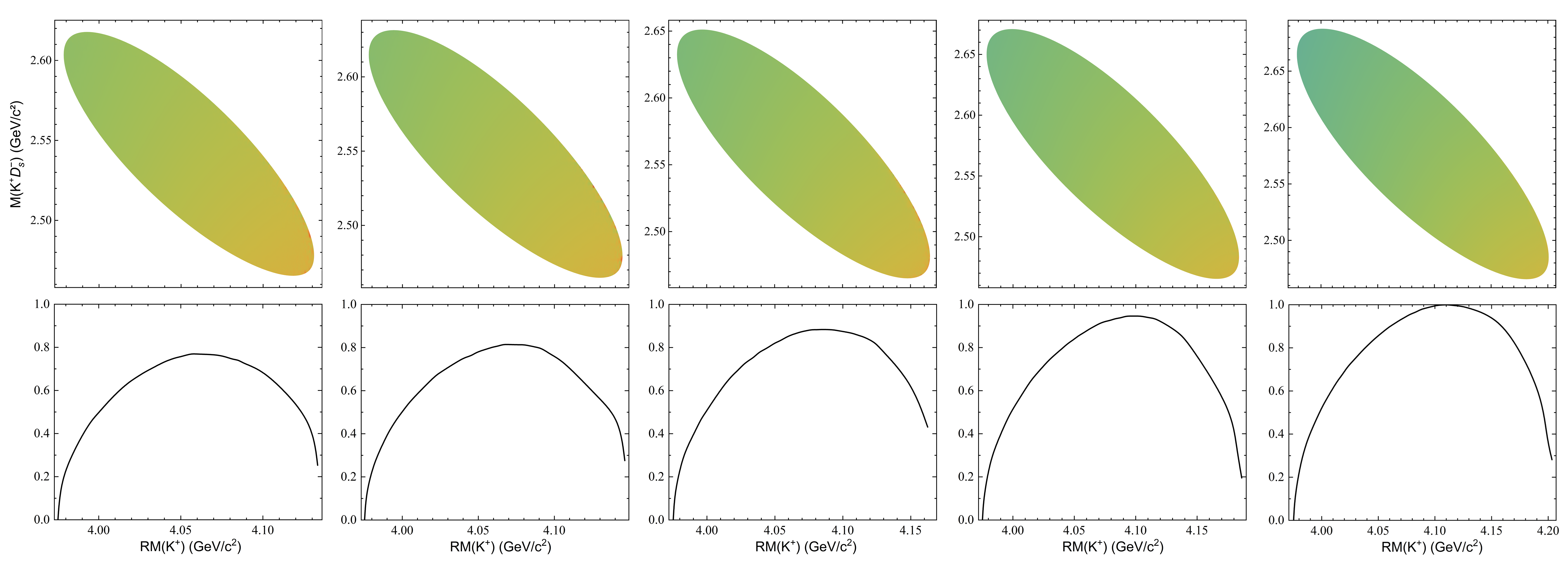}
	\caption{Dalitz plots and invariant mass spectra of the process $\psi \to D^-_s D^{*0} K^+$ in Fig.~\ref{fig:boxtu}(c). Further caption text as in \ref{fig:Invariantmassspectra1}.}
	\label{fig:Invariantmassspectra3}
\end{figure*}

\begin{figure*}
	\centering
	\includegraphics[width=0.95\linewidth]{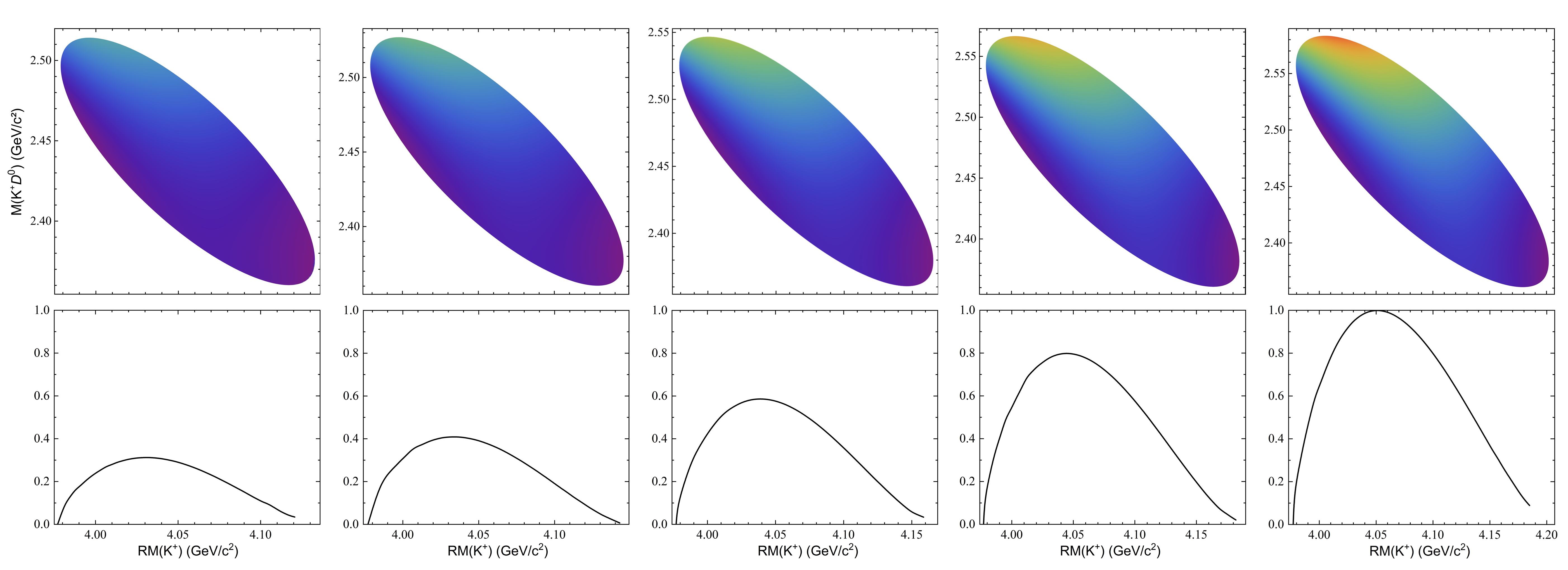}
	\caption{Dalitz plots and invariant mass spectra of the process $\psi \to D^{*-}_s D^0 K^+$ in Fig.~\ref{fig:boxtu}(d). Further caption text as in \ref{fig:Invariantmassspectra1}.}
	\label{fig:Invariantmassspectra4}
\end{figure*}

\begin{figure*}
	\centering
	\includegraphics[width=0.95\linewidth]{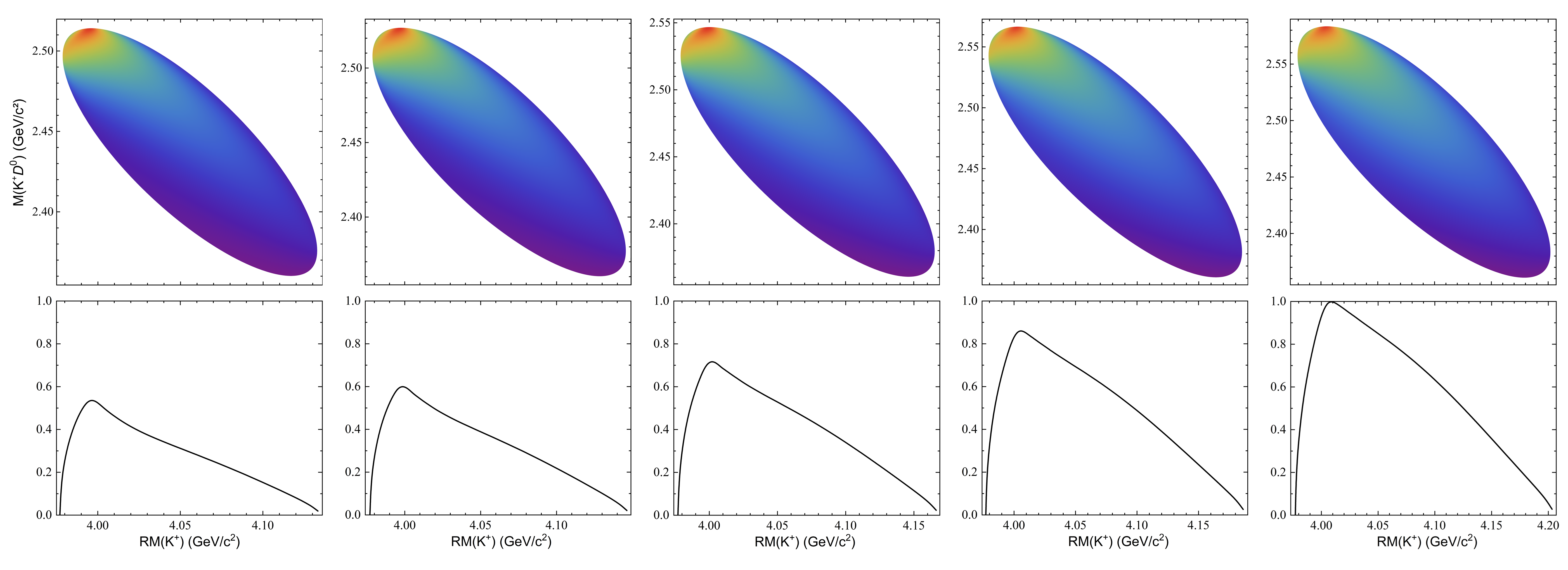}
	\caption{Dalitz plots and invariant mass spectra of the process $\psi \to D^{*-}_s D^0 K^+$ in Fig.~\ref{fig:boxtu}(e). Further caption text as in \ref{fig:Invariantmassspectra1}.}
	\label{fig:Invariantmassspectra5}
\end{figure*}

\begin{figure*}
	\centering
	\includegraphics[width=0.95\linewidth]{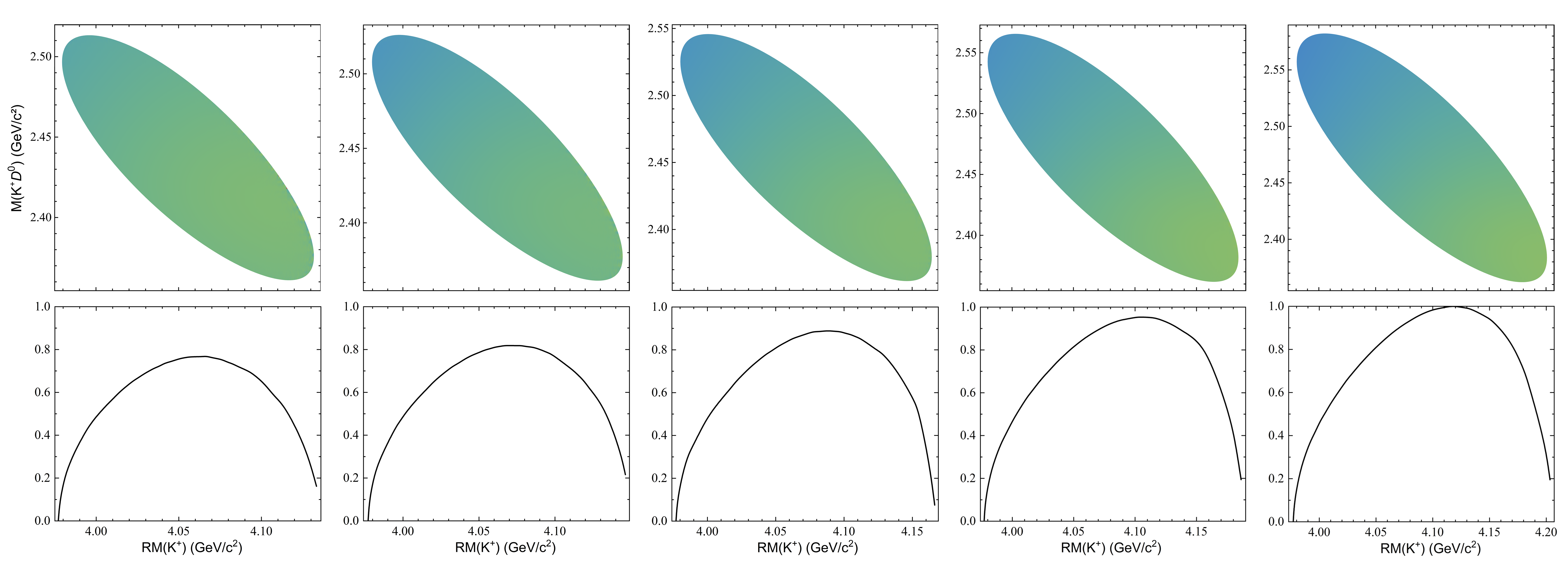}
	\caption{Dalitz plots and invariant mass spectra of the process $\psi \to D^{*-}_s D^0 K^+$ in Fig.~\ref{fig:boxtu}(f). Further caption text as in \ref{fig:Invariantmassspectra1}.}
	\label{fig:Invariantmassspectra6}
\end{figure*}

 
The box diagrams in Fig.~\ref{fig:boxtu} labeled (a) to (c) illustrate the process of $\psi \to D^-_s D^{*0} K^+$. 
In Figs.~\ref{fig:Invariantmassspectra1}, \ref{fig:Invariantmassspectra2} and \ref{fig:Invariantmassspectra3}, we present the corresponding Dalitz plots and their projections on the invariant mass $m_{23}$ (also referred to as $RM(K^+)$) for the decay $\psi \to D^-_s D^{*0} K^+$ at $\sqrt s = 4.628,\,4.641,\,4.661,\,4.681$, and $4.698$ GeV. 
Similarly, this approach also applies to the process for $\psi \to D^{*-}_s D^0 K^+$, for which the corresponding Dalitz plots and invariant mass spectra are depicted in Figs.~\ref{fig:Invariantmassspectra4}, \ref{fig:Invariantmassspectra5} and \ref{fig:Invariantmassspectra6}. 

Firstly, we examine the process $\psi \to Ds^- D^{*0} K^+$ presented in Figs.~\ref{fig:Invariantmassspectra1}, \ref{fig:Invariantmassspectra2}, and \ref{fig:Invariantmassspectra3}, being the results of the three Feynman diagrams in Fig.~\ref{fig:boxtu}(a-c), respectively. 
The contributions from these diagrams exhibit distinctly different linear structures. 
Notably, only Fig.~\ref{fig:Invariantmassspectra1} features a sharp peak around $3.985$ GeV in the high energy region of $\sqrt{s}$. 
This peak is particularly pronounced at $\sqrt{s} = 4.698$ GeV, and it disappears as the center-of-mass energy decreases, which is consistent with the kinematic characteristics of box singularities that arise only under specific energy conditions. 
It is noted that the threshold for the production of $D^*_{s2}(2573)^+ D^{*-}_s$ is 4685 MeV, indicating that the true box singularities can only occur when the center-of-mass energy exceeds this threshold. 
In Fig.~\ref{fig:Invariantmassspectra2}, a minor sharp peak structure is observed, with no significant variations in the energy region considered for $\sqrt{s}$. 
The primary difference from the previous diagram stems from the width of $D^*_1(2760)^0$, which is $177$ MeV, significantly broader than the resonant particle $D^*_{s2}(2573)$ of width $16.9$ MeV in Fig.~\ref{fig:Invariantmassspectra1}. 
Consequently, the box singularity effect in Fig.~\ref{fig:Invariantmassspectra2} is smoothed out. 
Conversely, the contribution of Fig.~\ref{fig:Invariantmassspectra3} is relatively featureless and lacks pronounced sharp structures; this is attributed in part to the wide nature of $\kappa(700)$, as well as the inability of the $D^0 D_s K^*_0$ vertex to guarantee all particles are on the mass shell, thereby preventing the occurrence of box singularities in this process. 
The lineshapes of these three box diagrams demonstrate significant differences, and the characteristics presented in the first and second diagrams correlate well with experimental findings, which we will analyze in further discussions.

Secondly, in the process $\psi \to D^{*-}_s D^0 K^+$, we observe markedly different line shapes and distributions. 
The discrepancies primarily originate from the contributions of the diagrams depicted in Fig.~\ref{fig:boxtu}(a) and Fig.~\ref{fig:boxtu}(d), where no sharp peak structure is present. 
This can be attributed to the fact that $D^*_1(2600)^0$ is a broad resonant state and, due to kinematic constraints, this diagram does not exhibit singularities, resulting in a smoother line shape. 
The remaining two Feynman diagrams associated with this process are consistent with those considered previously. 
Consequently, if the contributions to these two processes predominantly arise from the box diagrams, we should expect the two reactions to exhibit distinct Dalitz plots and invariant mass spectra. 

\begin{figure*}
	\centering
	\includegraphics[width=0.9\linewidth]{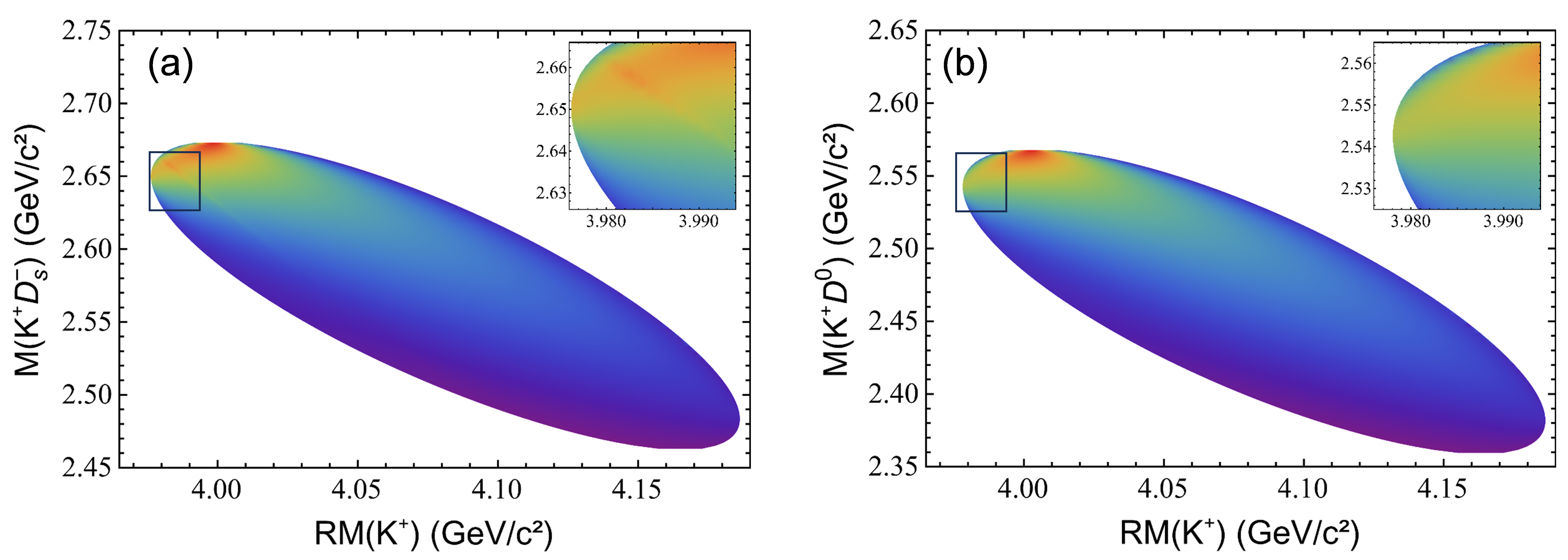}
	\caption{The Dalitz plots of $\psi \to D^-_s D^{*0} K^+$ (a) and $\psi \to D^{*-}_s D^0 K^+$ (b) under the same kind of total contributions. The center-of-mass energy $\sqrt s = 4.681$ GeV. The contribution of Fig.~\ref{fig:boxtu}(a) is the same as Fig.~\ref{fig:boxtu}(b) while Fig.~\ref{fig:boxtu}(d) and Fig.~\ref{fig:boxtu}(e) use equal proportions. The area marked by the blue box is enlarged as an inset in the top right corner to illustrate better the difference around 3.985 GeV.}
	\label{fig:Dalitzplotstotal}
\end{figure*}

\begin{figure}
	\centering
	\includegraphics[width=0.95\linewidth]{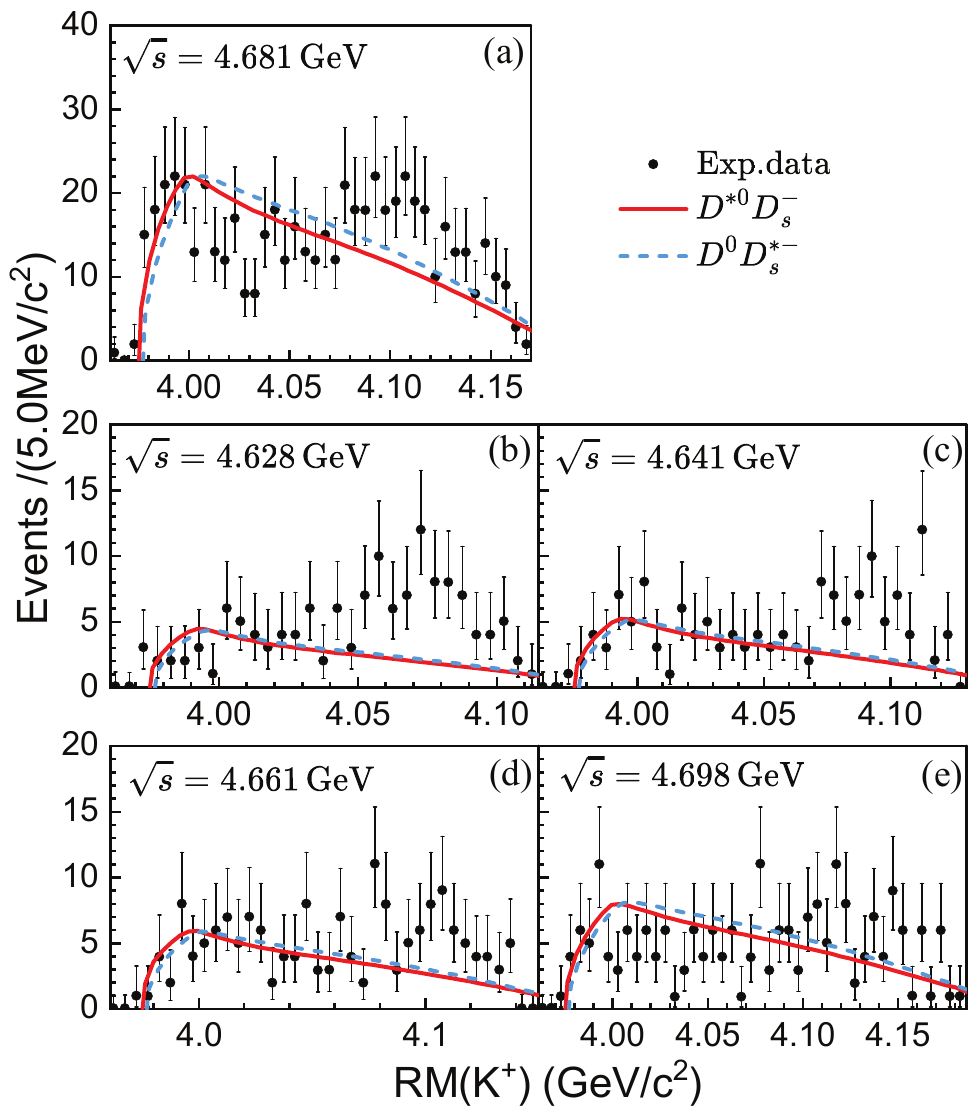}
	\caption{The invariant mass spectra of the processes $\psi \to D^-_s D^{*0} K^+$ and $\psi \to D^{*-}_s D^0 K^+$ under total contributions. The red solid line represents the results for $\psi \to D^-_s D^{*0} K^+$, while the blue dashed line describes the $\psi \to D^{*-}_s D^0 K^+$. The points with error bars are the experimental measurements taken from Ref.~\cite{BESIII:2020qkh}.}
	\label{fig:Invariantmassstotal}
\end{figure}

Thirdly, in Fig.~\ref{fig:Dalitzplotstotal}, we present the Dalitz plots for the three box diagrams corresponding to both reactions at a specified ratio. 
It is evident that within the energy region where the $Z_{cs}(3985)$ signal is observed, the two processes are indeed different. 
In other words, a significant contribution from $Z_{cs}(3985)$ is expected only in the invariant mass spectrum of $D^-_sD^{*0}$. 
Similarly, in Fig.~\ref{fig:Invariantmassstotal}, we show the invariant mass spectra for $D^-_sD^{*0}$ and $D^{*-}_s D^0$. 
Here, we consider the contribution of integrated luminosity. 
The integrated luminosity $L_{int}$ under five different energy points is listed in Ref.~\cite{BESIII:2020qkh}. 
Since we convince that the signal of $Z_{cs}(3985)$ is mainly from the box diagrams of $\psi \to D^-_s D^{*0} K^+$, we find the solid red lines are well consistent with the figures from experiment~\cite{BESIII:2020qkh}, especially around the threshold. 
We have also predicted the results of $\psi \to D^{*-}_s D^0 K^+$ using the same methods, which are illustrated with a blue dashed line.
Then we find that solid red line and the dashed blue line exhibit different behaviors at high $\sqrt{s}$.
Therefore, if experiments can effectively measure these two processes separately and obtain respective invariant mass spectra, it should clarify the role of the box diagrams in these reactions.

\subsection{ Discussion of box diagrams}

Through the calculations presented above, we have clearly demonstrated that contributions from the box diagrams can produce a pronounced peak structure near the threshold of the invariant mass of the $D^{-}_s D^{*0}$ system. 
These sharp peak structures may diminish with variations in energy, which aligns closely with experimental observations. 
Consequently, the box singularity is likely responsible for this behavior, suggesting that the observed peak corresponding to the resonance $Z_{cs}(3985)$ could be just the behavior of the loop singularity.
Therefore, a detailed investigation of the contributions from these box diagrams is essential to ascertain the genuine existence of $Z_{cs}(3985)$.

It is important to note that our analysis has focused solely on the contributions from the box diagrams, and thus our description for the high-energy region of the $D^{*0} D^{-}_s$ invariant mass spectrum may not be fully comprehensive. 
However, our calculations indicate that, in the vicinity of the threshold, the contributions from the box diagrams can adequately account for the existing experimental data. 
This highlights the necessity for more precise experimental measurements to clarify the significance of the contributions from the box diagrams.
On the other hand, we should 
admit that in the current model the ratios between these diagrams are estimated rather roughly, just relying on the current experimental data.

Furthermore, our calculations reveal that another final state, $D^0 D^{*-}_{s} K^+$, may exhibit distinct behavior near the threshold of  $D^0 D^{*-}_{s}$, warranting further experimental investigation.
Lastly, we posit that similar box diagram contributions may be relevant in a variety of other reaction processes, and this example should be carefully considered in future studies.

\section{Summary}\label{sec:summary}

In this work, we calculate several typical Feynman diagrams within box loop framework for two  processes of $\psi \to D^-_s D^{*0} K^+$ and $\psi \to D^{*-}_s D^0 K^+$, and provide the Dalitz plots and the invariant mass spectra at $\sqrt s = 4.628,\,4.641,\,4.661,\,4.681$, and $4.698$ GeV. 
These results open a new direction to understand the behavior at the thresholds of $D^-_s D^{*0}$ and $D^{*-}_s D^0$, where the signal of the $Z_{cs}(3985)$ resonance has been announced, particularly at $\sqrt s = 4.681$ GeV.
However, our calculations reveal that in the process $\psi \to D^-_s D^{*0} K^+$, the box loop diagram could produce the sharp peak structure around $3985$ MeV in the $ D^-_s D^{*0}$ invariant mass spectrum, whereas in the process $\psi \to D^{*-}_s D^0 K^+$, such a peak will disappear. 
Therefore, based on our calculations, it is possible that the signal of $Z_{cs}(3985)$ resonance near the threshold of $D^-_s D^{*0}$ and $D^{*-}_s D^0$ is indeed caused by box singularity rather than by the resonance state. 
This provides a potential box diagram explanations for $Z_{cs}(3985)$.
We should admit that current statistic is insufficient to draw definitive conclusions, thus, more experimental information is necessary.
Typically, we find that such box singularity contribution will be missing in the invariant mass spectrum of $D^{*-}_s D^0$. Therefore, it is urgent to make measurements for two  processes of $\psi \to D^-_s D^{*0} K^+$ and $\psi \to D^{*-}_s D^0 K^+$ separately.

\begin{acknowledgments}

We thank useful discussions and valuable comments from Yu Lu, Haojie Jing, Chaowei Shen, Shuming Wu, Fengkun Guo, and Bingsong Zou. 
This work is partly supported  by
the National Natural Science Foundation of China under Grant Nos. 12175239, 12221005, and 12475081; by the Chinese Academy of Sciences under Grant No. YSBR-101; by the Natural Science Foundation of Shandong Province under Grant No. ZR2022ZD26 and by Taishan Scholar Project of Shandong Province under Grant No. tsqn202103062.

\end{acknowledgments}

\bibliography{ref.bib}
\end{document}